\documentclass[aps,prd,12pt,superscriptaddress,eqsecnum,amsmath,nofootinbib,preprintnumbers]{revtex4-2}%

\usepackage{color,graphicx,float,subfigure}
\usepackage{amsfonts,amssymb,theorem,mathrsfs,times}
\usepackage{bm}
\usepackage{mathtools}
\usepackage{amsfonts,amssymb,theorem,mathrsfs}
\usepackage{dsfont}
\usepackage{setspace}
\usepackage{graphicx}
\usepackage{ulem}
\usepackage[colorlinks,linkcolor=blue,anchorcolor=blue,citecolor=blue]{hyperref}
\usepackage{orcidlink}
\usepackage{comment}

{\theorembodyfont{\upshape}
}
{\theorembodyfont{\upshape}
}
{\theorembodyfont{\upshape}
}
{\theorembodyfont{\upshape}
}
{\theorembodyfont{\upshape}
}
{\theorembodyfont{\upshape}
}

\newcommand{\dalm}{\kern1pt\vbox{\hrule height 0.9pt\hbox{\vrule width
0.9pt\hskip 2.5pt\vbox{\vskip 5.5pt}\hskip 3pt\vrule width
0.3pt}\hrule height 0.3pt}\kern1pt}

\begin{document}

\title{Images and photon regions of continuous photon sphere spacetime}

\author{Long-Yue Li \orcidlink{0000-0002-7785-1116}}
\email{lilongyue@nbu.edu.cn}
\affiliation{Institute of Fundamental Physics and Quantum Technology, Department of Physics, School of Physical Science and Technology, Ningbo University, Ningbo, Zhejiang 315211, China}

\author{ Xia-Yuan Liu \orcidlink{0009-0008-3852-0147}}
\email{liuxiayuan@mail.ustc.edu.cn}
\affiliation{
Interdisciplinary Center for Theoretical Study and Department of Modern Physics,\\
University of Science and Technology of China, Hefei, Anhui 230026,
China}

 \author{Rong-Gen Cai \orcidlink{0000-0002-3539-7103}}
 \email{caironggen@nbu.edu.cn}
\affiliation{Institute of Fundamental Physics and Quantum Technology,
 Department of Physics, School of Physical Science and Technology,
 Ningbo University, Ningbo, Zhejiang 315211, China}

\author{\\Yungui Gong \orcidlink{0000-0001-5065-2259}}
\email{gongyungui@nbu.edu.cn }
\affiliation{Institute of Fundamental Physics and Quantum Technology, Department of Physics, School of Physical Science and Technology, Ningbo University, Ningbo, Zhejiang 315211, China}

\author{Wenting Zhou \orcidlink{0000-0003-4046-753X}}
\email{zhouwenting@nbu.edu.cn}
\affiliation{Institute of Fundamental Physics and Quantum Technology, Department of Physics, School of Physical Science and Technology, Ningbo University, Ningbo, Zhejiang 315211, China}

\begin{abstract}
We study images of spacetimes containing continuous photon spheres (CPS).
For a self-gravitating, isotropic, spherically symmetric spacetime with CPS,
we find that a thin accretion disk produces images that closely resemble those of a Schwarzschild black hole, despite significant differences in photon dynamics.
More generally, for any static, spherically symmetric spacetime with a luminous CPS core, the image profile is universal:
members of this class produce identical image shapes, differing only by an overall normalization factor.
This universality is, however, sensitive to the nature of the accretion flow and breaks down for spherically symmetric infalling accretion,
where Doppler shifts and non-static emission introduce image features that depend on the flow dynamics and the metric.
Finally, we investigate photon regions in a rotating CPS spacetime and find that unlike in Kerr spacetime, the photon region appears as one or two angular sectors in a constant‑$\phi$ cross section.
These distinctive photon region properties could produce observable signatures that distinguish rotating CPS spacetimes from the Kerr one.
\end{abstract}

\maketitle

\section{Introduction}

A few years ago,
the Event Horizon Telescope (EHT) captured the first direct images of supermassive black holes at the center of M87~\cite{EventHorizonTelescope:2019dse} and Milky Way~\cite{EventHorizonTelescope:2022wkp}.
These groundbreaking observations provided compelling visual evidence for the existence of black holes and introduced a powerful new technique to study compact astrophysical objects through direct imaging.
The EHT images reveal a bright emission ring surrounding a dark shadow, formed by light bending around the black hole’s event horizon from the accretion disk.
A few decades ago, Luminet pioneered the visualization of a Schwarzschild black hole surrounded by a rotating thin accretion disk \cite{Luminet:1979nyg}.
Utilizing the ray-tracing method, researchers have since computationally reconstructed the appearance of black holes illuminated by their accretion flows \cite{Gralla:2019xty}.
By counting the number of times of light rays crossing the equatorial plane, the observed image can be classified into three components: direct emission (single crossing), the lensed ring (double crossings), and the photon ring (three or more crossings).
However, the lensed and photon rings lie extremely close to each other, and current telescope resolutions remain insufficient to resolve them distinctly.
Crucially, while the photon rings are governed primarily by the spacetime geometry and the properties of the photon sphere, direct emission depends sensitively on the spatial structure and emissivity profile of the accretion flow.
This interplay allows us to probe both the underlying spacetime metric and accretion physics through imaging observations.

Recent studies on various massive astrophysical objects—including black holes \cite{Peng:2020wun,Xu:2024gjs,Cao:2023par,Cao:2024kht,Cao:2024vtq,Sui:2023tje,Xiong:2025hjn,Meng:2023htc,Gan:2021xdl,Gan:2021pwu,Guo:2022ghl,Chen:2023qic,Zhang:2024lsf,Shen:2023nij,Zhang:2024jrw,Liang:2025bbn,Wang:2025dfn}, naked singularities \cite{Shaikh:2019hbm,Gyulchev:2020cvo,Gyulchev:2021dvt,Chen:2023knf,Deliyski:2024wmt,Wu:2024ixf,Shaikh:2018lcc,HassanPuttasiddappa:2025tji,Huang:2024bbs}, wormholes \cite{Huang:2023yqd,Chen:2024tss,Olmo:2021piq,Guerrero:2022qkh,Guo:2022iiy,Peng:2021osd,Luo:2023wru}, and other exotic compact objects \cite{Rosa:2022tfv,Huang:2025css,Tamm:2023wvn,Li:2025awg,Zhao:2025yhy,Guerrero:2022msp,Olmo:2023lil,Wang:2024uda}—have attracted considerable attention due to their extreme gravitational effects and potential to test fundamental physics.
Since black holes necessarily possess photon spheres \cite{Carballo-Rubio:2024uas,Cunha:2024ajc,Cunha:2020azh}, their images generally exhibit a characteristic photon ring.
Beyond the standard black hole images composed of direct emission, lensed ring, and photon ring, other intriguing scenarios have emerged.
For instance, the doppelganger black hole’s lensed and photon rings may be too narrow to resolve even in simulations, resulting in images resembling those of spacetimes without photon spheres \cite{Xu:2024gjs}.
In cases where the strong cosmic censorship conjecture is violated, multi-ring structures appear in the images \cite{Cao:2023par,Cao:2024kht}—a feature also common in images of naked singularities, wormholes, and other compact objects.
Additionally, black holes with multiple photon spheres, where an inner photon sphere possesses a higher effective potential, produce images with two photon rings or bright ring-like bands, rather than a single typical photon ring \cite{Meng:2023htc,Gan:2021xdl,Gan:2021pwu,Guo:2022ghl,Chen:2023qic}.
For naked singularities and certain compact objects that lack photon spheres, images consist solely of direct emission \cite{Huang:2024bbs}, whereas the presence of photon spheres can give rise to multi-ring structures \cite{Shaikh:2019hbm,Gyulchev:2020cvo,Gyulchev:2021dvt,Chen:2023knf,Deliyski:2024wmt,Huang:2024bbs,Wu:2024ixf}.
Wormhole images are particularly diverse, depending not only on the number of photon spheres but also on which universe the accretion disk and photon spheres occupy.
In more realistic settings, magnetohydrodynamic (MHD) simulations model the accretion flow around rotating black holes to generate their images \cite{Zhang:2024lsf,Shen:2023nij,Zhang:2024jrw}.
Astrophysical black holes are highly dynamic systems—evolving, actively accreting matter, and exhibiting complex variability—making the study of their time-dependent images crucial.
Recent efforts have begun tackling this challenge through time-dependent simulations \cite{Liang:2025bbn,Wang:2025dfn}.
Additionally, images of three black holes in static equilibrium configurations \cite{Li:2025jfq} and of axisymmetric but non-rotating spacetimes \cite{Chen:2025cwi} have also been explored.

The image of a massive compact object is intrinsically linked to the properties of its photon spheres.
Apart from black hole images, gravitational lensing is another observable phenomenon closely related to the photon sphere \cite{Tsukamoto:2021caq}.
In Schwarzschild spacetime, the photon sphere is an unstable circular orbit at radius $r=3M$, where $M$ is the black hole mass.
Other spherically symmetric spacetimes can admit multiple isolated photon spheres.
In axisymmetric but non-rotating spacetimes, the photon sphere becomes prolate on the meridional plane, losing perfect spherical symmetry \cite{Chen:2025cwi}.
When photon orbits become non-circular, non-closed, or when the spacetime is dynamic, the conventional notion of a photon sphere faces conceptual challenge.
Several attempts to generalize the definition have been proposed \cite{Claudel:2000yi,Amo:2023ofn}.
For example, in Boyer-Lindquist coordinates, the photon region of a Kerr black hole comprises unstable bound photon orbits oscillating latitudinally between fixed boundaries while maintaining a constant radius.
Similar photon region structures appear in other rotating spacetimes.
There has been substantial progress in the study of photon spheres \cite{Hod:2013jhd,Hod:2020pim,Hod:2012ax}.
In \cite{DiFilippo:2024poc}, the authors study the self-gravitating effect of a photon sphere .
They find that the mass of such a photon sphere modifies the external spacetime metric, which in turn can lead to the formation of additional photon spheres.
Furthermore, under certain conditions, this process may even produce a continuous distribution of photon sphere.
Notably, Hod recently investigated a class of spherically symmetric, self-gravitating isotropic (SSSGI) matter configurations in general relativity (GR) \cite{Hod:2025bsf}.
Intriguingly, this spacetime features an infinite continuum of photon spheres existing between radii $r_-$ and $r_+$.
Since photon sphere structure largely determines optical appearance, this configuration naturally raises fascinating questions about its observational image.
In this work, we present a systematic analysis of this remarkable problem.

This paper is organized as follows.
In Section \ref{S2}, we provide a brief introduction to the SSSGI spacetime with CPS.
Section \ref{S3} investigates its image formed by a thin accretion disk.
For comparison, Section \ref{S4} analyzes the image produced by spherically symmetric accretion.
Section \ref{S5} explores the construction of a rotating spacetime with CPS.
Finally, we summarize our conclusions and discuss future directions in Section \ref{S6}.

\section{Self-gravitating isotropic spherically symmetric spacetime with CPS}\label{S2}
In \cite{Hod:2025bsf}, the author investigated spherically symmetric, self-gravitating isotropic matter configurations within the framework of general relativity.
By solving the Einstein field equations, the study revealed the existence of objects with continuous photon spheres.
Furthermore, the metric parameter was constrained through the energy conditions.
In this section, we provide a concise overview of this work.

For a general static spherically symmetric spacetime, its metric can be expressed as
\begin{eqnarray}
ds^2=-f(r) dt^2+h^{-1}(r)dr^2+r^2d\Omega^2   \,,
\end{eqnarray}
where $d\Omega^2=d\theta^2+\sin^2\theta d\phi^2$ is the metric of two-dimensional sphere, and
\begin{eqnarray}\label{g}
h(r)=1-\frac{2m(r)}{r}  \,.
\end{eqnarray}

Consider a spacetime with CPS in the region $r_-\leqslant r\leqslant r_+$, and the matter in this region is described by the following energy-momentum tensor,
\begin{eqnarray}
T^\mu_\nu=\mathrm{diag}\left( -\rho, p, p, p \right)   \,.
\end{eqnarray}
For equatorial null geodesics ($\theta = \pi /2$), the null geodesic and relevant equation of motion for photons between $r_-$ and $r_+$ are
\begin{eqnarray}\label{nullgeo}
\dot{t}&=&\frac{E}{f}     \,,\nonumber\\
\dot{\phi}&=&\frac{L}{r^2}     \,,\nonumber\\
\dot{r}^2&=&h\left( \frac{E^2}{f}-\frac{L^2}{r^2} \right)      \,,\nonumber\\
\frac{d\phi}{dr}&=&\dfrac{1}{r^2\sqrt{\frac{1}{b^2}\frac{h}{f}-\frac{h}{r^2}}}     \,,
\end{eqnarray}
where $E$ and $L$ are energy and angular momentum of photons, a overdot `` $\cdot$ " represents the derivative with respect to affine parameter $\lambda$.
The photon sphere corresponds to the radius where $\dot{r}^2=(\dot{r}^2)'=0$, where `` $'$ " represents the derivative with respect to $r$.
This means
\begin{eqnarray}
r^2&=&fb^2    \,,\nonumber\\
\frac{2}{r^3}&=&\frac{f'}{f^2}\frac{1}{b^2}  \,,
\end{eqnarray}
or equivalently
\begin{eqnarray}\label{N}
N(r)=f'r-2f=0   \,,
\end{eqnarray}
where $b=L/E$ is the impact parameter.

The spacetime in the region $r_-\leqslant r\leqslant r_+$ contains continuous photon spheres.
That is, $N(r)=0$ for all $r_-\leqslant r\leqslant r_+$.
Therefore,
\begin{eqnarray}\label{f}
f=Cr^2     \,,
\end{eqnarray}
where $C$ is a positive constant.
As a result, for the spherically symmetric spacetime with CPS, the metric component takes the form $g_{tt}=-Cr^2$.
Substitute Eq. \eqref{f} into $\dot{r}^2=(\dot{r}^2)'=0$, we know that when the photons are located at CPS, they have the critical impact parameter
\begin{eqnarray}\label{bc}
b_c=\frac{1}{\sqrt{C}}  \,.
\end{eqnarray}

With this form of $g_{tt}$ established, the Einstein field equations now yield
\begin{eqnarray}
8\pi\label{rho} \rho&=&\frac{2m'}{r^2}      \,,\\
8\pi\label{p}  p&=&\frac{2(r-3m)}{r^3}=\frac{1-2m'}{r^2}   \,.
\end{eqnarray}
From Eq. \eqref{p}, we get the solution
\begin{eqnarray}
m(r)=\frac{1}{4} r+C_1 r^3  \,,
\end{eqnarray}
where $C_1$ is a constant.

Assuming the mass of the center core inside $r_-$ is $m_c$, and defining a dimensionless parameter $\mathcal{C}_c=m_c/r_-$, then
\begin{eqnarray}
r_-\mathcal{C}_c=m_c=m(r_-)=\frac{1}{4} r_-+C_1 r_-^3  \,.
\end{eqnarray}
Therefore,
\begin{eqnarray}
C_1=\frac{\mathcal{C}_c-\frac{1}{4}}{r_-^2}
\end{eqnarray}
and
\begin{eqnarray}
m(r)=\frac{1}{4} r+\frac{\mathcal{C}_c-\frac{1}{4}}{r_-^2} r^3  \,.
\end{eqnarray}

From Eqs. \eqref{g}, \eqref{rho} and \eqref{p}, we know
\begin{eqnarray}
\label{gr} h(r)&=&1-\frac{2m(r)}{r}=\frac{1}{2}-\frac{2}{r_-^2}(\mathcal{C}_c-\frac{1}{4})r^2  \,,\\
\label{rhor} \rho(r)&=&\frac{1}{16\pi r^2}+\frac{3}{4\pi r_-^2}(\mathcal{C}_c-\frac{1}{4})     \,,\\
\label{pr} p(r)&=&\frac{1}{16\pi r^2}-\frac{3}{4\pi r_-^2}(\mathcal{C}_c-\frac{1}{4})  \,.
\end{eqnarray}
Based on this metric, we can calculate the Ricci scalar and the Kretschmann scalar, and find that they are both divergent when $r$ tends to $0$.
Hence a CPS core extending to $r_- \rightarrow 0$ corresponds to a naked singularity.

Suppose that the outer edge of CPS, $r_+$, is also the outer edge of the self-gravitating matter, which means the region beyond $r_+$ is vacuum.
Then the pressure-less surface boundary condition can be written by
\begin{eqnarray}
p(r_+)=0   \,.
\end{eqnarray}
Therefore,
\begin{eqnarray}\label{rp}
r_+=\frac{r_-}{2}\frac{1}{\sqrt{3\left(\mathcal{C}_c-\frac{1}{4}\right)}}   \,,
\end{eqnarray}
and
\begin{eqnarray}
\mathcal{C}_c\geqslant\frac{1}{4}
\end{eqnarray}
Because of $r_+>r_-$,
\begin{eqnarray}
\mathcal{C}_c< \frac{1}{3}  \,.
\end{eqnarray}
The root of $h(r)=0$ is
\begin{eqnarray}
r_h=\frac{r_-}{2}(\mathcal{C}_c-\frac{1}{4})^{-\frac{1}{2}}>r_+  \,.
\end{eqnarray}
So there is no horizon within $ r_- \leqslant r \leqslant r_+$.

Assuming that the matter fields between $r_-$ and $r_+$ respect the dominant energy condition, $0\leqslant|p|\leqslant\rho$, then we get \cite{Hod:2025bsf}
\begin{eqnarray}
\mathcal{C}_c\geqslant \frac{1}{4}  \,.
\end{eqnarray}
From the strong energy condition, $\rho+3p\geqslant0$, we get
\begin{eqnarray}
\mathcal{C}_c\leqslant \frac{5}{12}  \,.
\end{eqnarray}
Combining them, then
\begin{eqnarray}\label{Cc}
\frac{1}{4}\leqslant\mathcal{C}_c< \frac{1}{3}  \,.
\end{eqnarray}

In this section, we obtain the spacetime with CPS via an alternative approach to Ref. \cite{Hod:2025bsf}.
We prove that the spherically symmetric spacetime with CPS must have a metric with component $g_{tt}=-Cr^2$.
Besides, filled with the self-gravitating isotropic matter fields in the region $r\in[r_-, r_+]$, $g_{rr}(r), p(r), \rho(r)$ and $m(r)$ can be obtained from the isotropic energy-momentum tensor and Einstein field equations.
The pressure-less outer edge determines $r_+$.
Finally, the metric and other physical quantities are determined by $r_-$ and $\mathcal{C}_c$.
From the energy conditions, the range of $\mathcal{C}_c$ is detremined.

\section{The image of the CPS naked singularity with thin accretion disk}\label{S3}
Studying the image of the CPS naked singularity is crucial for probing the nature of strong gravitational fields and light bending in its vicinity.
To analytically isolate these relativistic features, we employ a viewed face-on and optically and geometrically thin disk model, which reduces the complexity of accretion disk while retaining the essential effects.
Optically thin means negligible absorption and scattering, while geometrically thin implies negligible physical thickness.
In the SSSGI spacetime discussed in the previous section, the region outside $r_+$ is vacuum (i.e., the Schwarzschild spacetime).
We set $r_-\rightarrow 0$ and refer to this structure as the spacetime with a CPS core.

To ensure the metric components are continuous at $r_+$,
\begin{eqnarray}
f(r_+)=1-\frac{2M}{r_+}=h(r_+)  \,,
\end{eqnarray}
where $M$ is the mass parameter of the Schwarzschild spacetime.
Therefore, the metric of the whole spacetime can be written as
\begin{eqnarray}\label{metric}\label{metricy}
ds^2&=&-\frac{y-2}{y(yM)^2}r^2dt^2+\left( \frac{1}{2}-\frac{4-y}{2y(yM)^2} r^2 \right)^{-1}dr^2+r^2d\Omega^2  \qquad r<r_+    \,,\nonumber\\
ds^2&=&-\left(1-\frac{2M}{r}\right)dt^2+\left(1-\frac{2M}{r}\right)^{-1}dr^2+r^2d\Omega^2     \qquad r\geqslant r_+     \,,
\end{eqnarray}
where $r_+=yM$ and $y>2$ because $r_+$ must be located outside the horizon of the Schwarzschild spacetime.

If we adopt the pressure-less condition,
we can derive the following from \eqref{gr} and \eqref{rp}:
\begin{eqnarray}
h(r_+)=\frac{1}{3}  \,,
\end{eqnarray}
which means $y=3$.

We adopt an accretion disk model with radially decreasing emission, extending inward to the origin for the CPS spacetime, while to the horizon for Schwarzschild black hole.

The emitted intensity is \cite{Guerrero:2022qkh}
\begin{eqnarray}
I_{em}(r)=\dfrac{\pi/2-\arctan(r-2)}{\pi/2-\arctan(-2)}  \,,
\end{eqnarray}
where we have chosen $M=1$.

In the following section, we analyze the images of SSSGI spacetime with a thin accretion disk.
For the detailed methodology and the procedures used to derive these images, please refer to our previous works \cite{Cao:2023par,Cao:2024kht,Cao:2024vtq}.

\subsection{$y=3$}

\begin{figure}[htb]
    \centering
    \includegraphics[width=6cm]{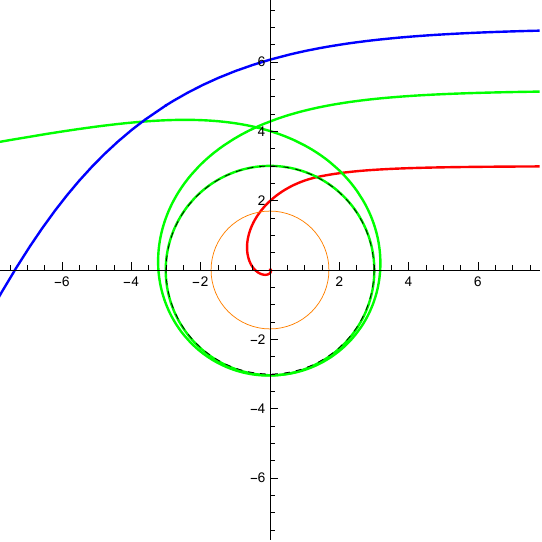}
    \caption{The ray trajectory with $r_+=3M$.
    The dashed circle is $r=3M$.
    The green curve and blue curve represent rays with $b>b_c$.
    The orange curve represents the ray with $b=b_c$.
    The red curve represents the ray with $b<b_c$. }
    \label{traCPSrp3M}
\end{figure}

For the SSSGI spacetime, $r_+=3M$, and $b_c=3\sqrt{3}M=b_{c\,Sch}$, where $b_{c\,Sch}$ is the critical impact parameter of Schwarzschild spacetime.
The pressure at $r_+$ is vashing, i.e., $p(r_+)=0$.
After perturbation, the photons at circular orbits $r<3M$ will shift to neighboring orbits, preserving their circular trajectory.
While at $r=3M$, the  perturbed photons will fly to infinity or remain in circular motion on slightly closer orbits.
The ray trajectory is shown in Fig. \ref{traCPSrp3M}.
When $b>b_c$ (as shown by the green and blue curves in Fig. \ref{traCPSrp3M}), the rays begin and end at infinity, and remain entirely within the outer (Schwarzschild) spacetime throughout its trajectory.
When $b=b_c$ (as shown by the orange curve), the ray is located at CPS.
It is worth noting that because of continuous photon spheres, $b=b_c$ can be at any arbitrary $r\leqslant r_+$. The orange curve in the figure is just one of the situations arbitrarily selected for illustration.
When $b<b_c$ (as shown by the red curve), the ray falls into the singularity.

\begin{figure}[htbp]
		\begin{minipage}[t]{0.45\linewidth}
			\centering
			\includegraphics[height=4cm]{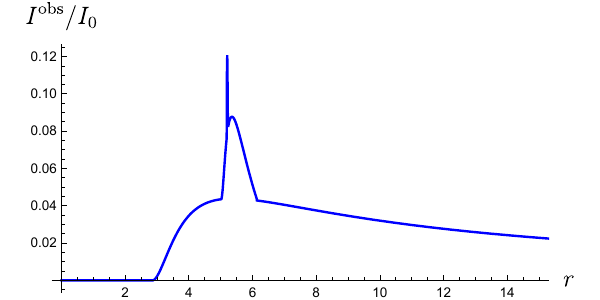}
		\end{minipage}%
		\begin{minipage}[t]{0.45\linewidth}
			\centering
			\includegraphics[height=4cm]{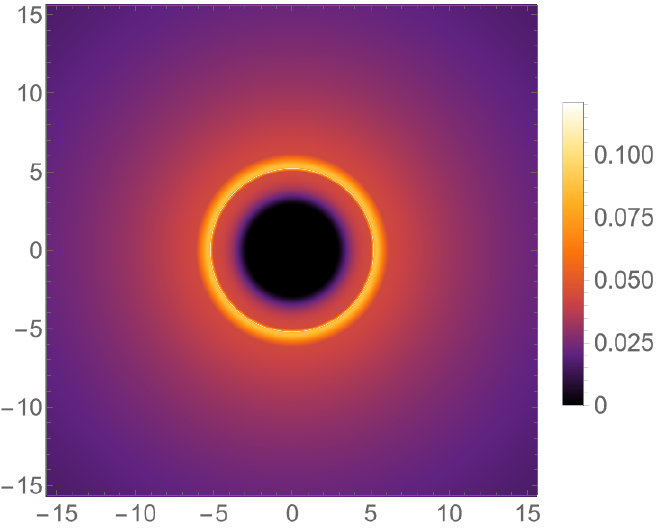}
		\end{minipage}%

		\begin{minipage}[t]{0.45\linewidth}
			\centering
			\includegraphics[height=4cm]{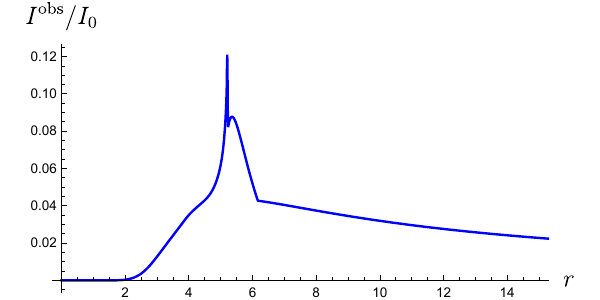}
		\end{minipage}%
		\begin{minipage}[t]{0.45\linewidth}
			\centering
			\includegraphics[height=4cm]{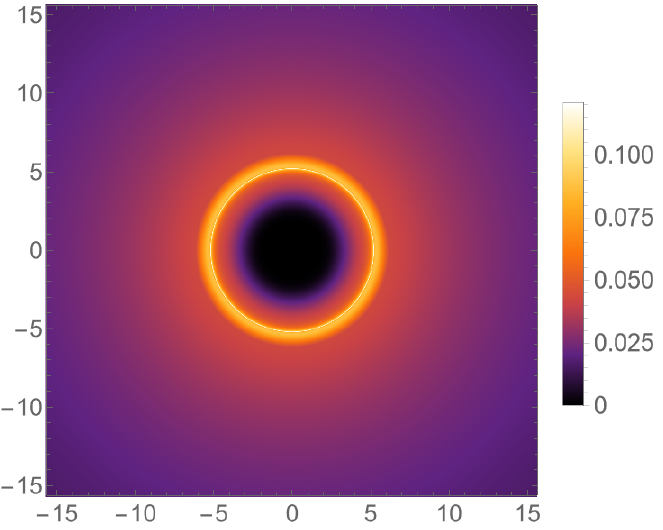}
		\end{minipage}%

  \caption{The observed intensities (left column) and images (right column) of Schwarzschild black hole (top) and the SSSGI naked singularity with a CPS core for $y=3$ (bottom).
  $I_0$ is the maximum emitted intensity.}
  \label{imager3M}
\end{figure}

The observed intensities and images of Schwarzschild black hole and the SSSGI spacetime with a CPS core for $y=3$ are shown in Fig. \ref{imager3M}.
These two images are remarkably similar because the ray trajectories with $b>b_c$ are identical in both spacetimes, leading to the same observed intensities.
However, there are subtle  differences in the $b<b_c$ regime.
For instance, the case of Schwarzschild spacetime exhibits a slightly larger shadow radius, which might allow us to distinguish them, though this minor differentiation  makes such discrimination extremely challenging in practice.

Although the images of Schwarzschild black hole and the SSSGI spacetime with a CPS core both have shadow, their causes are different.
For Schwarzschild black hole, the rays with small $b$ end at the horizon without passing through the accretion disk.
Consequently, no ray reaches the center of the image, resulting in a dark area.
However, for SSSGI spacetime with a CPS core,
the scenario is distinct.
The photons with small $b$, e.g. $b < b_c$, orbit numerous times near the center, undergoing multiple intersections with the accretion disk before finally reaching the observer.
However, these rays experience extreme redshift, resulting in significantly darkness in the center of the image.

\subsection{$y=3.5$}

$r_+$ is closely related to the critical impact parameter $b_c$, which affects the photon sphere.
Since we intend to study the properties of the photon sphere, we want $b_c$ to be able to take a wide range of values.
To achieve this, we relax the pressure-less surface boundary condition $p(r_+)=0$ in the following subsections, thereby allowing $r_+$ to vary freely.
The dominant energy condition and strong energy condition result in
\begin{eqnarray}
\frac{1}{4}\leqslant \mathcal{C}_c\leqslant \frac{5}{12} \,,\nonumber\\
r_+ \leqslant \frac{r_-}{2\sqrt{3(\mathcal{C}_c-\frac{1}{4})}}   \,.
\end{eqnarray}
In the general formulation, $r_-$ and $\mathcal{C}_c$ are independent constants, with $\mathcal{C}_c$ capable of approaching $1/4$.
Hence we can choose suitable $\mathcal{C}_c$ based on some stellar model.
For an object with density $\rho \sim 1/r^2$ (as Eq. \eqref{rhor}), the mass scales as $m \sim r$, implying $\mathcal{C}_c=m_-/r_- \sim \text{const}$.
For an object with constant density, $m\sim r^3$.
Therefore, a natural choice is $\mathcal{C}_c = A^2 r_-^2 + 1/4$, where $A > 0$ is a constant.
Substituting this into the expression for $r_+$ yields the bound $r_+ \leqslant 1/(2\sqrt{3}A)$.
Consequently, by choosing a suitable $A$, the upper bound on $r_+$ can be made arbitrarily large, allowing $r_+$ to take any value.

Inside $r_+$, all circular photon orbits share the same critical impact parameter,
\begin{eqnarray}
b_c(y)=yM\sqrt{\frac{y}{y-2}}  \,.
\end{eqnarray}
which attains its minimum $b_{c, min}=3\sqrt{3}M$ at $y = 3$.

In the case of $y=3.5$, $b_c= 5.35M>5.20M= b_{c\,Sch} $.
The ray trajectory is similar to the case of $y=3$.
For the rays begin at infinity, they will fly to infinity for $b>b_c$, fall into the singularity for $b<b_c$, and perpetually orbit at $r\leqslant3.5M$ for $b=b_c$.

\begin{figure}[htbp]
		\begin{minipage}[t]{0.45\linewidth}
			\centering
			\includegraphics[height=4cm]{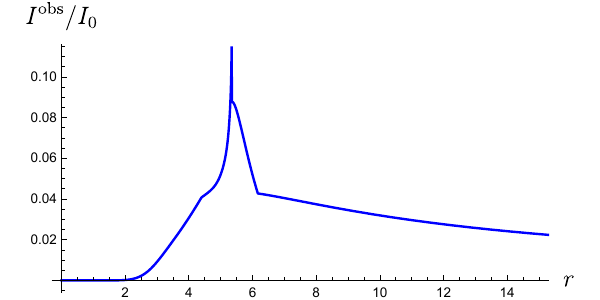}
		\end{minipage}%
		\begin{minipage}[t]{0.45\linewidth}
			\centering
			\includegraphics[height=4cm]{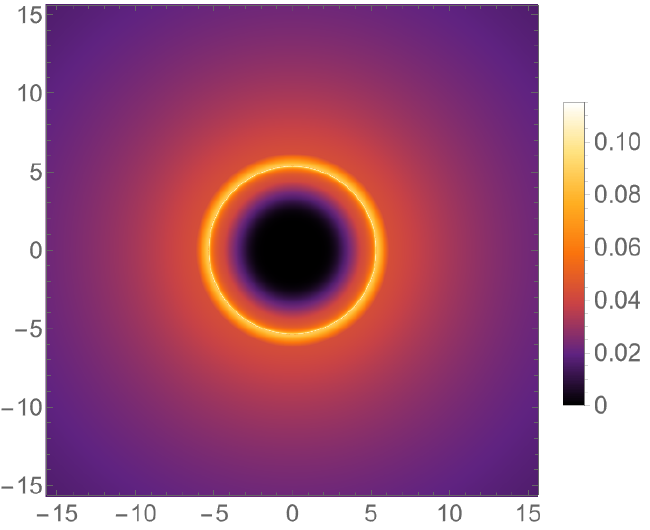}
		\end{minipage}%
  \caption{The observed intensity (left) and image (right) of the SSSGI with a CPS core for $y=3.5$.}
  \label{imager3p5M}
\end{figure}

The observed intensity and image of the SSSGI spacetime with a CPS core for $y=3.5$ are shown in Fig. \ref{imager3p5M}.
Although the photon ring is a little larger than the Schwarzschild's one, it is hard to distinguish them because their photon rings are too close.

\subsection{$y=2.5$}
In the case of $y=2.5$, $b_c= 5.59M> b_{c\,Sch} $.
Therefore, there are two impact parameters, $5.59M$ corresponds to $r\leqslant2.5M$ and $5.20M$ corresponds to $r=3M$.
The ray trajectories for different $b_c$ are more complex.
\begin{itemize}
\item
For $b>5.59M$, the photons begin from infinity, pass through the perihelion, travel outward, and eventually return to infinity.

\item
For $b=5.59M$, the photons are located at the CPS for any $r\leqslant2.5M$.

\item
For $5.20M<b<5.59M$, there are two kinds of rays.
One begins and ends at infinity, and they have a perihelion larger than $3M$.
Another begins and ends at the singularity, and they have a aphelion between $2.5M$ and $3M$.
Obviously, the latter can not be received by observer at infinity.
Therefore, they has no effect on observations.

\item
For $b=5.20M$, the photons are located at the isolated photon sphere $r=3M$.

\item
For $b<5.20M$, the photons monotonically fall from infinity into the singularity, or vice versa.

\end{itemize}

\begin{figure}[htb]
    \centering
    \includegraphics[width=6cm]{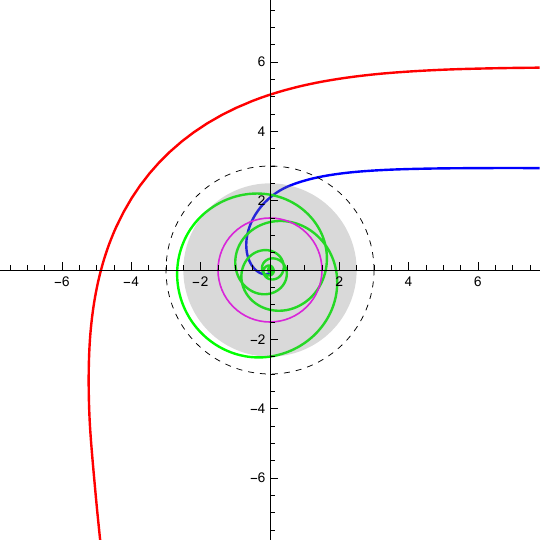}
    \caption{The ray trajectory with $r_+=2.5M$.
    The dashed circle is $r=3M$ corresponding to the photon sphere with $b = b_{c, Sch} = 5.20M$.
    It is an isolated photon sphere.
    The gray region is the CPS ($r\leqslant2.5M$). Every circle in this region, such as the magenta curve, is a photon sphere.
    The red curve represents the ray with $b>5.20M$.
    The green curve represents the ray with $5.20M< b<5.59M$.
    The blue curve represents the ray with $b<b_c$. }
    \label{tra2p5}
\end{figure}

These different kinds of rays trajectories are shown in Fig. \ref{tra2p5}.

\begin{figure}[htbp]
		\begin{minipage}[t]{0.45\linewidth}
			\centering
			\includegraphics[height=4cm]{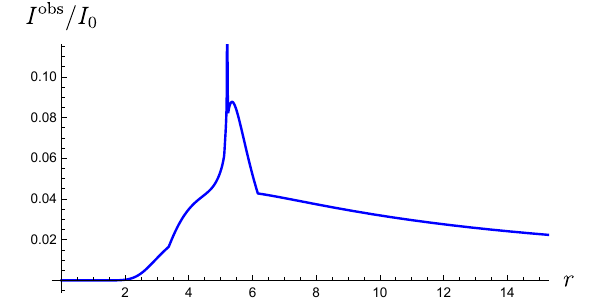}
		\end{minipage}%
		\begin{minipage}[t]{0.45\linewidth}
			\centering
			\includegraphics[height=4cm]{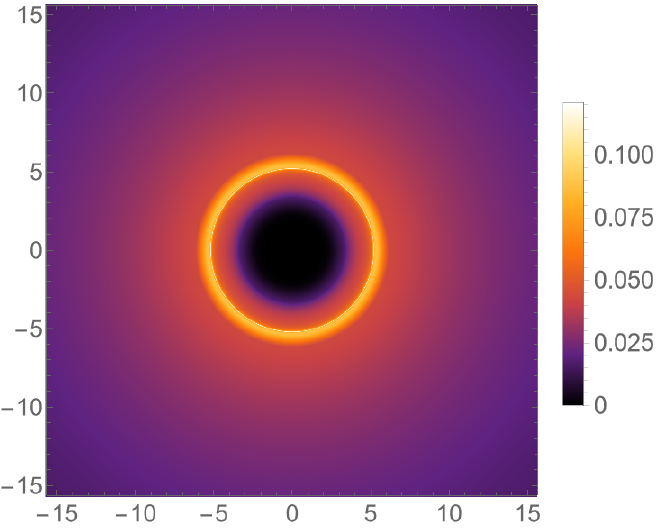}
		\end{minipage}%
  \caption{The observed intensity (left) and image (right) of the SSSGI naked singularity with a CPS core for $y=2.5$.}
  \label{imager2p5M}
\end{figure}

The observed intensity and image of the SSSGI spacetime with a CPS core for $y=2.5$ are shown in Fig. \ref{imager2p5M}.
The image is similar to the Schwarzschild's one, and their photon rings are located at identical positions.

In this section, we study the images of SSSGI spacetime with a CPS core for different $r_+$.
Although the behavior of the rays of the spacetime with these three kinds of $r_+$ are very different (including deviations from Schwarzschild's behavior), they generate similar observable image features.
This suggests that $r_+$ has negligible impact on the observational appearance, despite (i) different $r_+$ corresponding to distinct critical impact parameters $b_c$, (ii) $r_+$ determining the existence of isolated photon sphere, and (iii) certain $r_+$ even producing two separate $b_c$.
In order to identify characteristic observational features of the spacetime with a CPS core, we will study its image with a spherically symmetric accretion disk in next section.

\section{Image of CPS core with spherical accretion}\label{S4}

\subsection{Spherically symmetric spacetime with a CPS core}
In section \ref{S2}, we studied the metric of the spacetime with CPS, and found that the metric component takes the form $g_{tt}=-Cr^2$.
This result is independent of the underlying gravitational theory and the energy-momentum tensor of source.
As a matter of fact, there is always a singularity in spherically symmetric spacetime with a CPS core distributed down to $r=0$.
The proof is as follows.

The metric of the static spherically symmetric spacetime with CPS can be written as
\begin{eqnarray}\label{metric}
ds^2&=&-Cr^2 dt^2+\frac{1}{h(r)}dr^2+r^2d\Omega^2        \,,\nonumber\\
h(r)&=&1-\frac{2m(r)}{r}   \,.
\end{eqnarray}
The corresponding Ricci scalar and Kretschmann scalar are
\begin{eqnarray}
\label{R}   R&=&\frac{6m+6rm'-4r}{r^3}    \,,\\
\label{K}   K&=&R^{\mu\nu\lambda\tau}R_{\mu\nu\lambda\tau}=\frac{60m^2-8mr(4+3m')+4r^2(2+3m'^2)}{r^6}    \,.
\end{eqnarray}
Assuming $R$ equals $R_0$ when $r=0$, then
\begin{eqnarray}
6(mr)'= r(4+r^2R_0)\rightarrow 4r\,,   \qquad \text{when } r\rightarrow0 \,.
\end{eqnarray}
Therefore,
\begin{eqnarray}\label{mr0}
m\rightarrow \frac{r}{3}+\frac{C_2}{r}  \,,   \qquad \text{when } r\rightarrow0 \,,
\end{eqnarray}
where $C_2$ is a constant.
Substituting Eq. \eqref{mr0} into Eq. \eqref{K},
\begin{eqnarray}
K\rightarrow\frac{8(36C_2^2+r^4)}{3r^8}\,,   \qquad \text{when }  r\rightarrow0 \,.
\end{eqnarray}
Hence $K$ diverges when $r$ tends to $0$.
As a result, $R$ and $K$ can not both remain finite at $r=0$. Therefore,
the spherically symmetric spacetime with a CPS core extending to $r=0$ must have a singularity.

\subsection{The image of CPS spacetime with static spherically symmetric accretion}
We consider a general static, spherically symmetric spacetime that contains a CPS core and in which matter is distributed only inside $r_+$, while the exterior ($r>r_+$) is vacuum.
The accreting matter within the core emits radiation, so the observed image is produced by photon trajectories that traverse the interior and exterior regions and encode the spacetime geometry.
This setup therefore defines a static, spherically symmetric accretion model.
For the methodology and detailed techniques used to compute images of spherically symmetric accretion, see e.g. \cite{Xiong:2025hjn,Chen:2025ifv,Meng:2024puu,He:2021htq,Qin:2020xzu,Zare:2024dtf}.
Below we provide a brief overview.

For the thick accretion, the observed intensity $I_{obs}$ is determined by integrating the specific emissivity along the photon path\footnote{For monochromatic light, $I_\nu\sim g_i^3$. But after integrating over $\nu$, $I=\int I_\nu d\nu\sim g_i^4$.}
\begin{eqnarray}\label{Iobs}
I_{obs}=\int j(r) \cdot g_i^4(r) \cdot dl  \,,
\end{eqnarray}
where $j(r)$ represents the emissivity per unit volume and is usually given by $j(r)\propto1/r^2$,
$g_i=\sqrt{f(r_{em})/f(r_{obs})}$ is the redshift factor,
$r_{em}$ and $r_{obs}$ are the positions for static luminous matter and observer, respectively,
and $l$ represents the proper distance along the photon's path,
\begin{eqnarray}\label{l1}
dl_{static}=\sqrt{\frac{1}{h(r)}dr^2+r^2d\phi^2}=\sqrt{\frac{1}{h(r)}+r^2\left( \frac{d\phi}{dr} \right)^2}dr  \,.
\end{eqnarray}
For the CPS spacetime, substituting Eqs. \eqref{nullgeo} and \eqref{f} into Eq. \eqref{l1}, we get
\begin{eqnarray}
\label{l}
dl_{static}=\sqrt{\frac{1}{h}+r^2 \frac{1}{r^4}\left(\frac{1}{b^2}\frac{h}{Cr^2}-\frac{h}{r^2}\right)^{-1}}dr = \frac{1}{\sqrt{h(1-b^2C)}}dr \,.
\end{eqnarray}
Therefore, using Eq. \eqref{bc}, the observed intensity is
\begin{eqnarray}\label{Iobssta}
I_{obs}=\frac{1}{\sqrt{1-b^2C}}\int_0^{r_+} \frac{j(r) \cdot g_i^4(r)}{\sqrt{h(r)}} \cdot dr \propto \dfrac{1}{\sqrt{1-b^2/b_c^2}} \,.
\end{eqnarray}

\begin{figure}[htbp]
		\begin{minipage}[t]{0.45\linewidth}
			\centering
			\includegraphics[height=4cm]{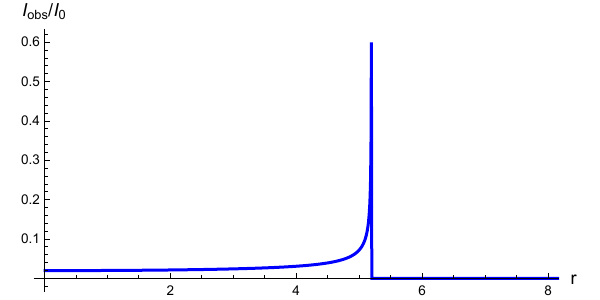}
		\end{minipage}%
		\begin{minipage}[t]{0.45\linewidth}
			\centering
			\includegraphics[height=4cm]{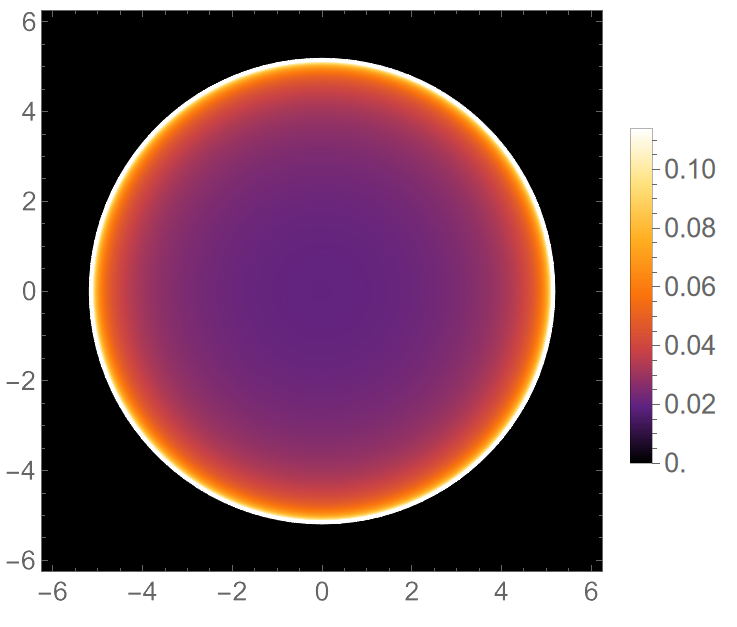}
		\end{minipage}%

		\begin{minipage}[t]{0.45\linewidth}
			\centering
			\includegraphics[height=4cm]{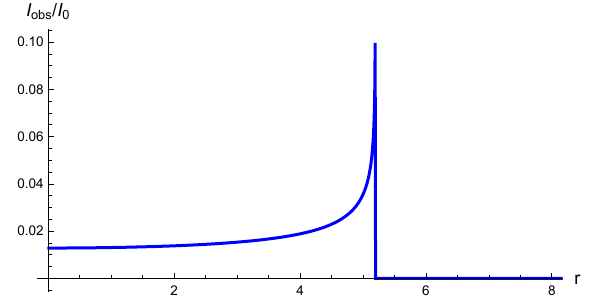}
		\end{minipage}%
		\begin{minipage}[t]{0.45\linewidth}
			\centering
			\includegraphics[height=4cm]{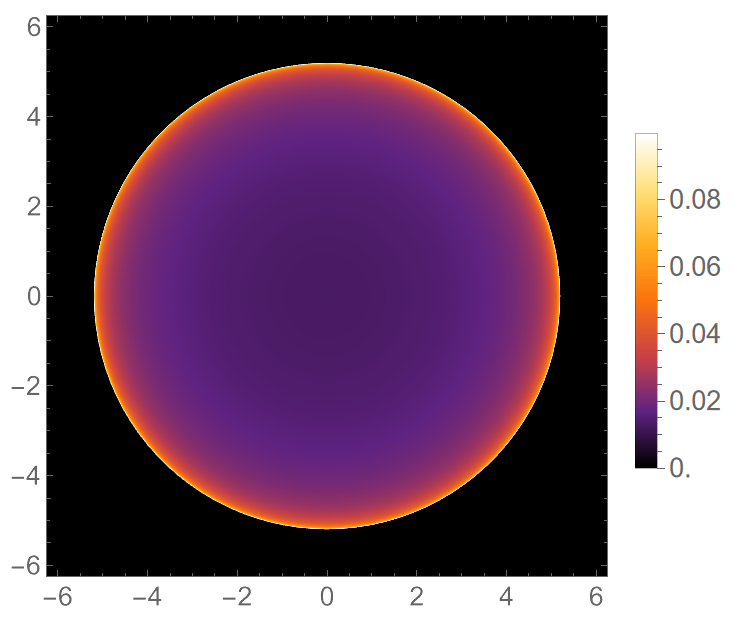}
		\end{minipage}%

  \caption{The observed intensities (left column) and images (right column) of SSSGI spacetime (top) and Schwarzschild black hole (bottom) with static spherically symmetric accretion.
  The luminous material is only distributed within $r_+=3M$ for CPS spacetime and between $2M$ and $3M$ for Schwarzschild black hole.}
  \label{ImageCPSSta}
\end{figure}

The observed intensity for other compact objects depends on both the luminosity model and the spacetime metric.
For example, for a Schwarzschild black hole one finds
\begin{eqnarray}
I_{obs}=\int_{2M}^{3M} j(r) \cdot g_i^4(r) \sqrt{\frac{1}{1-\frac{2M}{r}}+r^2\frac{1}{r^4}\left(\frac{1}{b^2}-\frac{1-2M/r}{r^2}\right)^{-1}}dr  \,.
\end{eqnarray}
Here the integration region corresponds to the radiating region in the chosen model.
Fig. \ref{ImageCPSSta} compares the image of a CPS spacetime with $r_+ = 3M$ to that of the Schwarzschild spacetime for the same static, spherically symmetric accretion model.
The two shadows have the same size, but the CPS spacetime produces a brighter observed intensity.
More generally, while the intensity profile of many compact objects depends on both the emissivity model $j(r)$ and the metric functions,
a remarkable simplification occurs for any spherically symmetric spacetime with a static luminous CPS core.
In that case the dependence on the impact parameter $b$ factorizes entirely into the universal prefactor $(1-b^2/b_c^2)^{-1/2}$ (see Eq. \eqref{Iobssta}): the $b$-dependence is independent of $g_{rr}$ and of the detailed radial profile $j(r)$.
The information about the matter distribution and redshift only affects the overall amplitude $\int j(r) g_i^4(r)/\sqrt{h(r)} dr$.
Consequently, the image profiles for this class of spacetimes are identical except an overall normalization.
This conclusion is robust: it does not depend on the particular gravitational theory, the detailed form of the energy–momentum tensor of source, or the emitted intensity profile.
The critical impact parameter $b_c$, however, is determined by the coefficient in $g_{tt}$ and therefore does depend on the underlying theory and source properties;
$j(r)$ and $h(r)$ only set the overall brightness. This property could therefore aid in identifying stellar objects that contain a CPS core in astronomical data.
At the same time, there are important limitations.
Because the image profile depends on a single parameter $b_c$ and the emissivity affects only the normalization,
it is difficult to reconstruct the underlying gravitational theory,
the detailed energy–momentum tensor of source, or the emission distribution from the image alone.

For the image of the CPS spacetime with parameter $ r_+ = 3M $ in Fig. \ref{ImageCPSSta},
we also include a comparison with the image of the Schwarzschild spacetime under a static spherically symmetric accretion.
We can observe that the shadows of the CPS spacetime and the Schwarzschild spacetime are the same size,
but the CPS spacetime has a brighter observed intensity.

\subsection{The image of CPS spacetime with infalling spherically symmetric accretion}
Another commonly used and simple spherically symmetric accretion models is the infalling accretion.
In this case, the background spacetime is composed of non-luminous matter, and the free-falling, radiating matter does not affect the background metric.
The projection length $dl$ of $dx^\mu$ onto the observer's four-velocity is
\begin{equation}
dl=-\frac{1}{g_i}\left.\left( K_t \frac{1}{\sqrt{f}} \right)\right|_o \frac{1}{K^r}dr,
\end{equation}
where $K^\mu$ is the photon four-momentum,
the corresponding redshift factor $g_i$ is
\begin{equation}
g_i=\frac{K_\mu u_o^\mu}{K_\mu u_e^\mu}=\frac{f}{ 1+b\sqrt{(1-f)(1/b^2-f/r^2)}},
\end{equation}
the subscripts $o$ and $e$ denote evaluation at the observer location $r_{obs}$ and the emission location $r_{em}$, respectively.
Here we have taken the observer to be at infinity.
Note that $g_i$ does not depend on $g_{rr}$;
rather, $dl$ depends on the redshift factor $g_i$ and on the components of $K^\mu$, both of which depend on the impact parameter $b$.
Physically, this reflects the fact that rays passing through the same spacetime point but arriving from different directions (i.e., having different $b$) undergo different Doppler shifts.
Consequently, $g_i$ depends on $b$, so $b$ cannot be factored out of the integral in Eq. \eqref{Iobs}.
As a result, the simplifying properties of Eq. \eqref{Iobssta} are not present for this infalling model.
Therefore, for the infalling spherically symmetric accretion model, $I_{obs}$ depends on the particular emission model, and hence on the source energy–momentum tensor and other factors,
just as it does in other spacetimes.

\begin{figure}[htbp]
		\begin{minipage}[t]{0.45\linewidth}
			\centering
			\includegraphics[height=4cm]{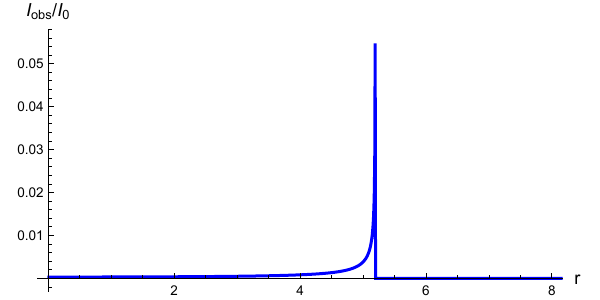}
		\end{minipage}%
		\begin{minipage}[t]{0.45\linewidth}
			\centering
			\includegraphics[height=4cm]{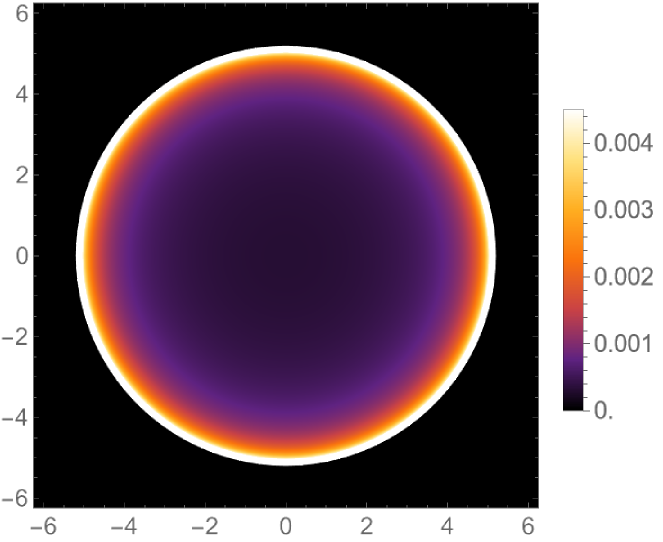}
		\end{minipage}%

		\begin{minipage}[t]{0.45\linewidth}
			\centering
			\includegraphics[height=4cm]{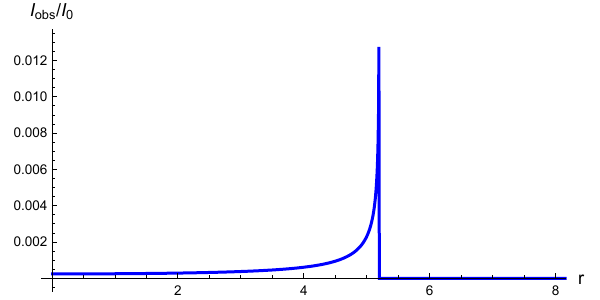}
		\end{minipage}%
		\begin{minipage}[t]{0.45\linewidth}
			\centering
			\includegraphics[height=4cm]{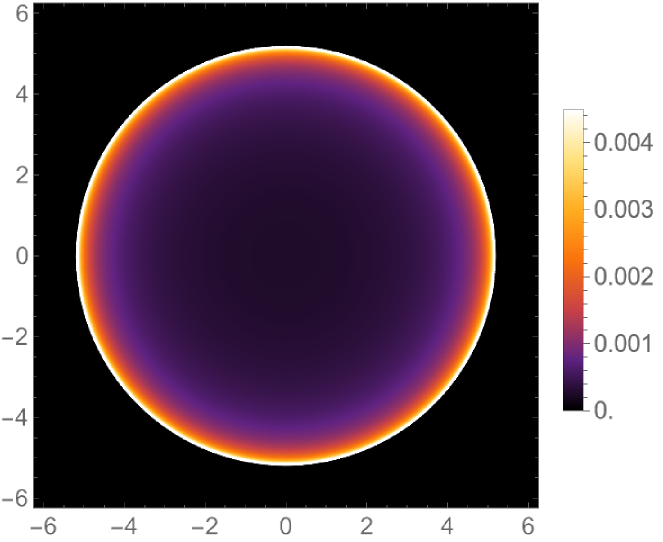}
		\end{minipage}%

  \caption{The observed intensities (left column) and images (right column) of CPS spacetime (top) and Schwarzschild black hole (bottom) with infalling spherically symmetric accretion.
  The luminous material is only distributed within $r_+=3M$ for CPS spacetime and between $2M$ and $3M$ for Schwarzschild black hole.}
  \label{ImageCPSFall}
\end{figure}

Figure \ref{ImageCPSFall} shows the image of CPS spacetime with infalling spherically symmetric accretion.
The metric used is given in Eq. \ref{metricy} with $y = 3$.
As in the static spherically symmetric accretion case,
the CPS spacetime and the Schwarzschild black hole produce shadows of the same apparent size;
however, the Schwarzschild image appears dimmer.

\section{Photon region of rotating CPS spacetime}\label{S5}

\begin{figure}[htb]
    \centering
    \includegraphics[width=4cm]{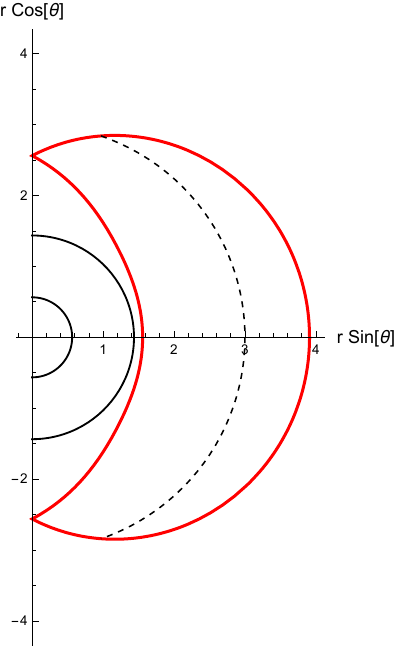}
    \caption{The photon region of Kerr black hole ($a=0.95M$) viewed in a cross-section along a constant-$\phi$ plane.
    The black curves represent the event horizon and Cauchy horizon.
    The dashed curve represents a photon orbit in the photon region.}
    \label{PSKerrBH}
\end{figure}

\begin{figure}[htbp]
		\begin{minipage}[t]{0.45\linewidth}
			\centering
			\includegraphics[height=6cm]{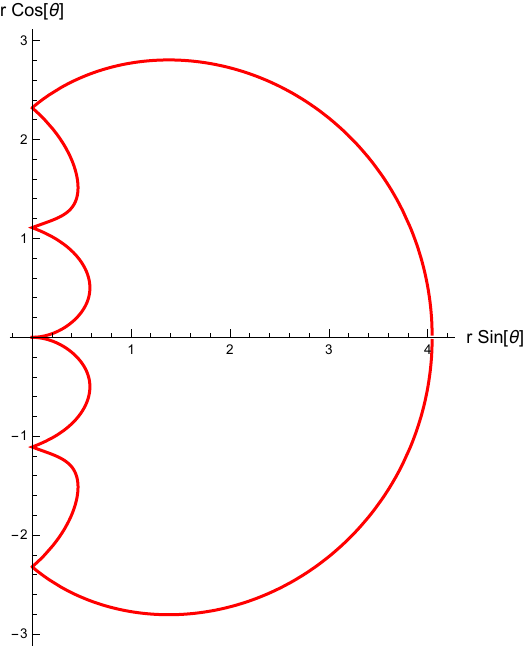}
		\end{minipage}%
		\begin{minipage}[t]{0.45\linewidth}
			\centering
			\includegraphics[height=6cm]{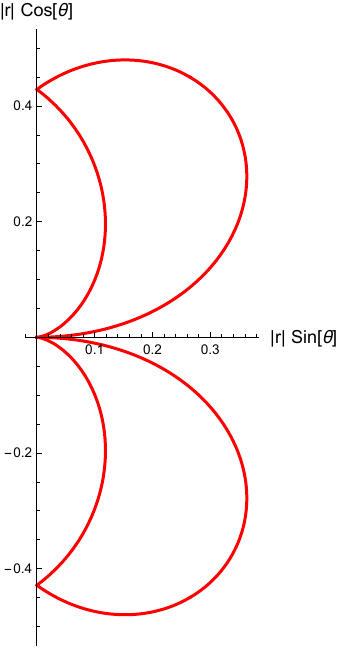}
		\end{minipage}%
  \caption{The photon region of Kerr naked singularity ($a=1.05M$) viewed in a cross-section along a constant-$\phi$ plane.
  The left diagram  is for $r>0$ and the right one is for $r<0$.}
  \label{PSKerrNS}
\end{figure}

Many spherically symmetric spacetime possess one or several photon spheres,
where photons can orbit on closed circular trajectories,
balancing between falling inward and escaping to infinity.
However, in axisymmetric spacetime, the shape of photon spheres (more generally, the photon region) is various.
For example, in the Boyer-Lindquist coordinate the photon region of a Kerr black hole consists of unstable bound orbits whose rays oscillate in latitude ($\theta$) between fixed limits at a constant radial coordinate ($r$).
When viewed in cross-section along a constant-$\phi$ plane, the Kerr photon region appears as a crescent as shown in Fig. \ref{PSKerrBH};
the photon region of a Kerr naked singularity is shown in Fig. \ref{PSKerrNS}.
For another example, a Schwarzschild black holes immersed in Born-Infeld magnetic fields, which is axisymmetric but non-rotating, has a circular equatorial photon sphere that becomes prolate on the meridional plane,
showing elongation along the polar axis \cite{Chen:2025cwi}.
In this section, we study the photon sphere for a rotating CPS spacetime.

Suppose a CPS core exists inside $r_+$ in a stationary, rotating spacetime with the metric,
\begin{eqnarray}\label{rometric}
ds^2&=&-\left( 1-\frac{r^2+a^2-\Delta}{\rho^2} \right)dt^2+\frac{\rho^2}{\Delta}dr^2-\frac{2a\sin^2\theta}{\rho^2}(r^2+a^2-\Delta)dtd\phi
+\rho^2d\theta^2+\frac{\Sigma\sin^2\theta}{\rho^2}d\phi^2    \,,\nonumber\\
\rho^2&=&r^2+a^2\cos^2\theta    \,,\nonumber\\
\Sigma&=&(r^2+a^2)^2-a^2\Delta\sin^2\theta    \,,
\end{eqnarray}
where $\Delta(r)$ is to be determined and $a$ is the rotation parameter.
The null geodesics in this CPS spacetime are
\begin{eqnarray}
\label{nullt}\rho^2\dot{t}&=&a(L-Ea\sin^2\theta)+\frac{1}{\Delta}(r^2+a^2)[(r^2+a^2)E-aL]    \,,\\
\label{nullphi}\rho^2\dot{\phi}&=&\frac{L-Ea\sin^2\theta}{\sin^2\theta}+\frac{a}{\Delta}[(r^2+a^2)E-aL]    \,,\\
\label{nulltheta}\rho^4\dot{\theta}^2&=&K-\frac{(L-Ea\sin^2\theta)^2}{\sin^2\theta}\equiv\Theta(\theta)    \,,\\
\label{nullr}\rho^4\dot{r}^2&=&[(r^2+a^2)E-aL]^2-\Delta K\equiv R(r)  \,,
\end{eqnarray}
where $K$ is the Carter constant.
For convenience,
we define the impact parameter $b$ = $L/E$ and the reduced Carter constant $K_E=K/E^2$.

The ``photon sphere" corresponds to null orbits satisfying $R(r)=R'(r)=0$.
If there is a CPS core inside $r_+$, then this condition holds for all $r<r_+$, so
\begin{equation}
\Delta(r)=\frac{(r^2+a^2-ab)^2}{K_E}.
\end{equation}
Because the metric should be invariant (up to the sign of $g_{t\phi}$) under $a\rightarrow -a$, it is convenient to rewrite this as
\begin{eqnarray}
\label{delta}
\Delta(r)=\frac{(r^2+a^2-B)^2}{k},
\end{eqnarray}
so that photons on the CPS orbits have impact parameter $b=B/a$ and reduced Carter constant $K_E=k$.
Here  \( a \), \( B \), and \( k \) are spacetime parameters,
while \( b \) and \( K_E \) label photon trajectories.
The Ricci scalar and Kretschmann scalar diverge at $r=0$, $\theta=\pi/2$,
so a ring singularity exists at the center, as in Kerr spacetime.
Moreover, $\Delta\geqslant0$ for all $r<r_+$, so no regular (nondegenerate) horizon is present.

From Eq. \eqref{nulltheta}, the null geodesic in the $\theta$-direction takes the form
\begin{eqnarray}\label{thetadot}
\rho^4\sin^2\theta\cdot\dot{\theta}^2=-a^2\sin^4\theta+(k+2B)\sin^2\theta-b^2   \,.
\end{eqnarray}
Since the left-hand side of the above equation is always positive,
$\theta$ must lie within ranges determined by the roots of the right-hand side.
Notably, these limits are independent of $r$,
unlike the Kerr case where the allowed latitude depends on $r$.
Defining $x=\sin^2\theta$ with $0<x<1$,
the right-hand side of Eq. \eqref{thetadot} becomes
\begin{eqnarray}
A(x)=-a^2x^2+(k+2B)x-b^2.
\end{eqnarray}
Two qualitatively distinct photon-region types arise:
(i)Photons oscillate within the polar angular range $(\theta_0, \pi-\theta_0)$.
(ii)Photons are confined to two separate latitudinal zones $(\theta_1, \theta_2)\cup(\pi-\theta_2, \pi-\theta_1)$.

For case (i), $A(\sin\theta_0)=0$ and $A(\sin (\pi/2))\geqslant0$.
In other words, $A(0)<0$ and $A(1)\geqslant0$.
This reduces to
\begin{eqnarray}
k>(a-B/a)^2\geqslant0    \,,
\end{eqnarray}
and $\theta_0$ is determined by
\begin{eqnarray}
\sin^2\theta_0=\frac{k+2B-\sqrt{k(k+4B)}}{2a^2}  \,.
\end{eqnarray}

\begin{figure}[htb]
		\begin{minipage}[t]{0.45\linewidth}
			\centering
			\includegraphics[height=6cm]{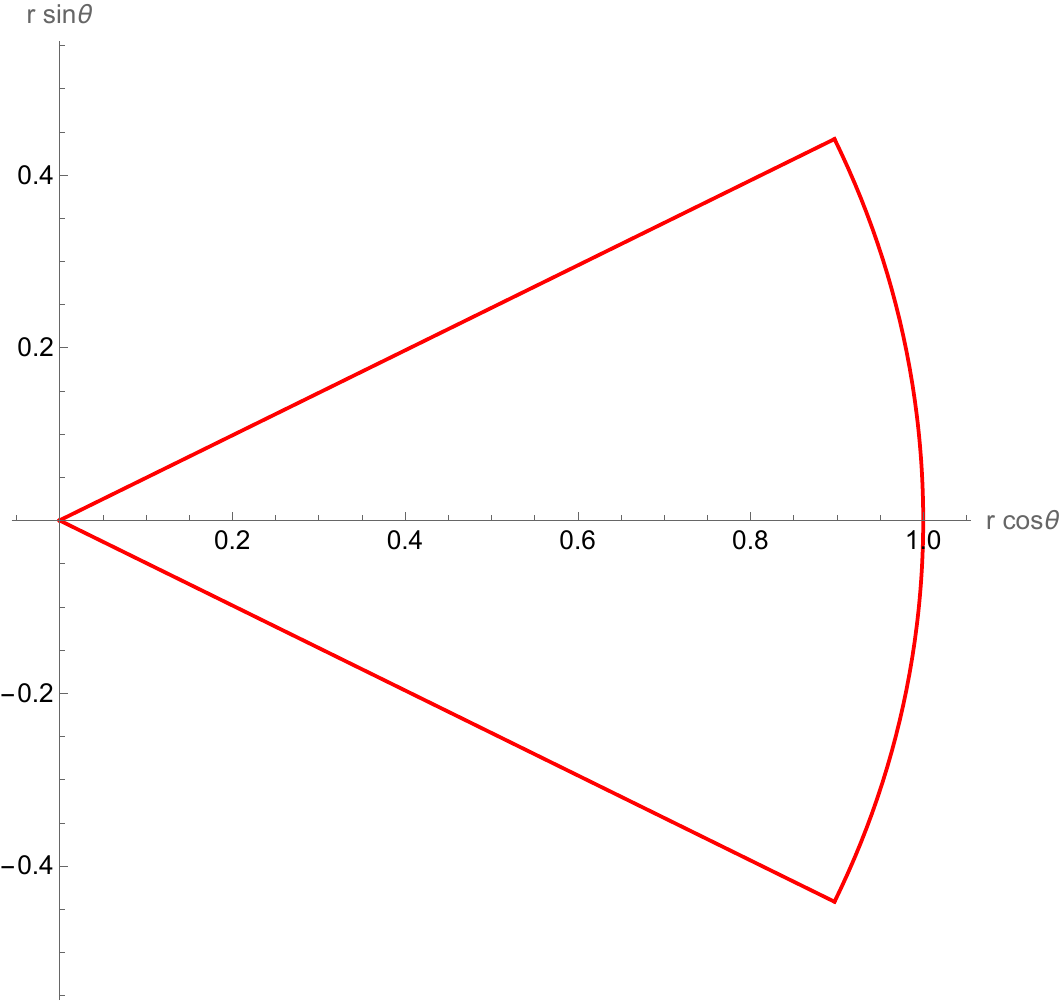}
		\end{minipage}%
		\begin{minipage}[t]{0.45\linewidth}
			\centering
			\includegraphics[height=6cm]{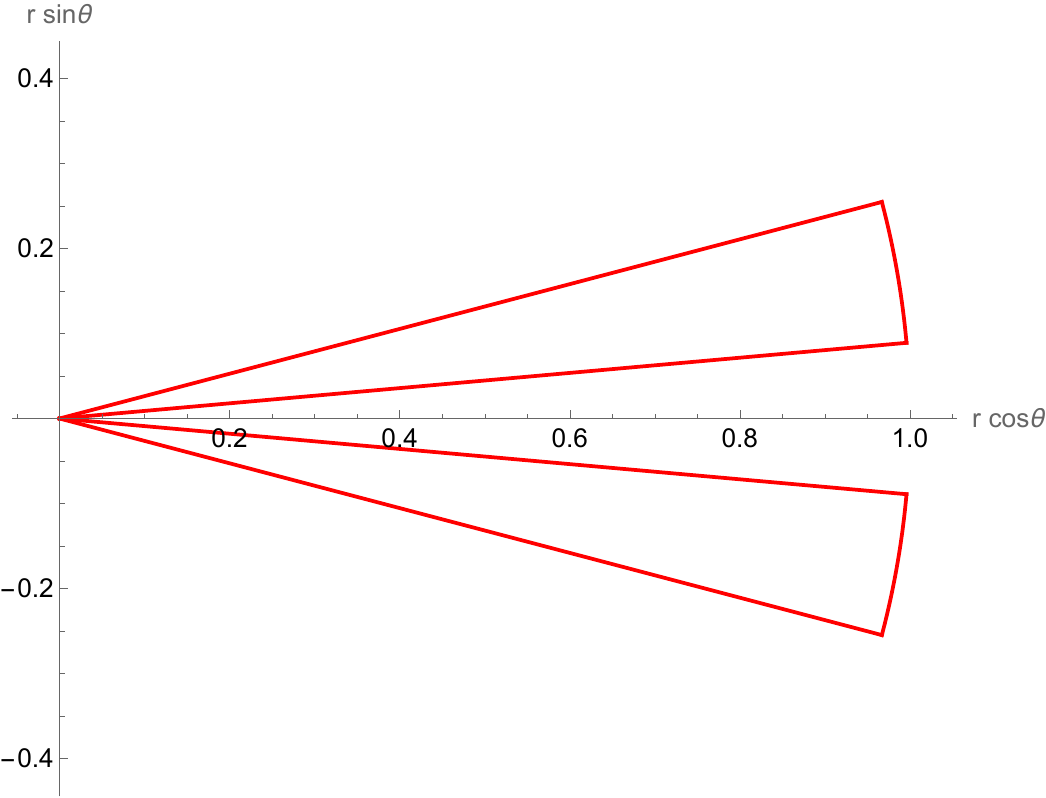}
		\end{minipage}%
    \caption{The schematic diagram of the photon region of the rotating CPS spacetime viewed in a cross-section along a constant-$\phi$ plane.
    The left figure is for case (i) with spacetime parameters $a=0.5, B=0.1$ and $k=1$.
    The right one is for case (ii) with spacetime parameters $a=0.5, B=0.1$ and $k=0.05$.}
    \label{roCPSps}
\end{figure}

For case (ii), $A(\sin\theta_1)=A(\sin\theta_2)=0$, and $0<\theta_1<\theta_2<\pi/2$.
This means $A(0)<0$, $A(1)<0$ and the maximum of $A(x)$ in $(0,1)$ is positive.
These conditions can be written as
\begin{equation}
\begin{split}
0<\frac{k+2B}{2a^2}<1,\\
4B^2-(k+2B)^2<0,\\
k<a^2+b^2-2B,
\end{split}
\end{equation}
which simplify to the compact bound,
\begin{eqnarray}
\max(-2B, -4B, 0)<k<\min(2a^2-2B, a^2+(B/a)^2-2B).
\end{eqnarray}
Correspondingly, $\theta_1$ and $\theta_2$ are determined by
\begin{equation}
\begin{split}
\sin^2\theta_1= \frac{k+2B - \sqrt{k(k+4B)}}{2a^2},\\
\sin^2\theta_2= \frac{k+2B + \sqrt{k(k+4B)}}{2a^2}.
\end{split}
\end{equation}
A schematic cross-section of the photon region at constant $\phi$ is shown in Fig. \ref{roCPSps}.
This structure differs from Kerr case in several important ways.
First, in the rotating CPS spacetime different constant-$r$ photon orbits share identical $\theta$-limits;
all unstable photon orbits have the same impact parameter $b$ and reduced Carter constant $K_E$, determined entirely by the spacetime parameters $a$, $k$, and $B$. In Kerr spacetime $b$ and $K_E$ vary with $r$.
Second, under perturbations photons on the unstable Kerr orbits typically either fall into the singularity or escape to infinity.
In the rotating CPS spacetime, aside from some photons at $r_+$ that can escape to infinity, most perturbed photons migrate to adjacent unstable orbits and continue to orbit.
These distinctive features of the rotating CPS photon region provide potentially observable signatures that can distinguish this spacetime from Kerr: differences may appear in shadows, images, and gravitational lensing.
With improving observational capabilities, detection of such rotating CPS objects may be feasible in the foreseeable future.

\section{Conclusion and discussion}
\label{S6}

In this paper we investigated images of spacetimes containing CPS,
focusing on a simplified treatment of the spherically symmetric, self‑gravitating isotropic (SSSGI) spacetime.
This solution describes a naked singularity whose interior ($r < r_+$) contains CPS while the exterior is the Schwarzschild vacuum;
in any spherically symmetric spacetime that hosts a CPS core the $tt-$component of the metric is found to take the form $g_{tt}=-Cr^2$;
energy conditions require the exterior matching to satisfy $2M < r_+ \leqslant 4M$.
When $3M \leqslant  r_+ \leqslant 4M$ the photon spheres form a continuous distribution sharing a single impact parameter,
whereas for $2M < r_+ < 3M$ the photon sphere structure includes both a continuous band and a discrete sphere at $r = 3M$,
yielding two distinct impact parameters.
Remarkably, for all allowed $r_+$ photons emitted arbitrarily close to the singularity can escape to infinity and be observed,
in stark contrast to the Schwarzschild black hole.

Nevertheless, images produced by a thin accretion disk around the SSSGI naked singularity closely resemble those of Schwarzschild because photons emitted very near the singularity are highly redshifted and contribute little to the observed intensity.
Besides, the multi‑ring features seen in some naked singularity with thin‑disk models are absent here.
For a static, spherically symmetric luminous core surrounded by vacuum we showed that the observed intensity has a universal $b$‑dependence (Eq. \eqref{Iobssta}), independent of the detailed form of $g_{rr}$,
the emissivity profile, or even the gravitational theory,
although the overall normalization and the critical impact parameter $b_c$ depend on the spacetime and source.
If instead the luminous matter is infalling and the background is non‑luminous, this universality breaks down because rays emitted from the same spatial point in different directions have different impact parameters and redshifts,
so the observed intensity depends on the emission model and metric details.

Motivated by Kerr black holes, we proposed a rotating CPS ansatz (Eqs. \eqref{rometric} and \eqref{delta}) and found that all CPS photon orbits share the same impact parameter $b = B/a$ and reduced Carter constant $K_E = k$;
their $\theta$‑limits are radius‑independent, so the photon region in a constant‑$\phi$ slice appears as one or two angular sectors,
a structure distinct from Kerr spacetime and potentially producing observable differences in shadows and lensing.

Finally, we stress that the rotating metric presented is a mathematical construction without a demonstrated physical interior matching to Kerr:
it does not satisfy the standard requirements for a physically viable Kerr interior (smooth matching, vanishing surface pressure, energy conditions, and the correct non‑rotating limit),
so constructing a consistent rotating interior containing CPS remains an open problem for future work.

\section*{Acknowledgement}
This research is supported in part by the National Natural Science Foundation of China key project under Grant No. 12535002.



%

\end{document}